\documentclass{nature}

\usepackage{lettrine}
\usepackage{graphicx}
\usepackage{dcolumn}
\usepackage{bm}
\usepackage{amssymb}
\usepackage{amssymb}
\usepackage{epsfig}
\usepackage[10pt]{moresize}
\usepackage{comment}
\usepackage{graphicx}
\usepackage{epsfig}
\usepackage{subfigure}

\title{Fast and simple scheme for generating NOON states of photons in circuit QED}
\author{Qi-Ping Su $^{1}$, Chui-Ping Yang $^{1,3}$, and Shi-Biao Zheng $^{2\star}$}
\begin{document}

\maketitle
\begin{affiliations}
\item Department of Physics, Hangzhou Normal University,
Hangzhou, Zhejiang 310036, China
\item Department of Physics, Fuzhou University, Fuzhou 350002,
China
\item State Key Laboratory of Precision Spectroscopy,
Department of Physics, East China Normal University, Shanghai
200062, China
\\$^\star$e-mail:sbzheng11@163.com
\end{affiliations}

\begin{abstract}
The generation, manipulation and fundamental understanding of entanglement lies at
very heart of quantum mechanics. Among various types of entangled states, the
NOON states are a kind of special quantum entangled states with two orthogonal component
states in maximal superposition, which have a wide range of potential applications in quantum communication
and quantum information processing. Here, we propose a fast and simple scheme for generating NOON states of photons in
two superconducting resonators by using a single superconducting transmon
qutrit. Because only one superconducting qutrit and two resonators
are used, the experimental setup for this scheme is much simplified when
compared with the previous proposals requiring a setup of two
superconducting qutrits and three cavities. In addition, this scheme is
easier and faster to implement than the previous proposals, which require
using a complex microwave pulse, or a small pulse Rabi frequency in order to
avoid nonresonant transitions.
\end{abstract}


\lettrine[lines=2]{V}arious physical systems have been considered for
building up quantum information processors. Among them, circuit QED
consisting of microwave resonators and superconducting qubits is
particularly appealing [1,2]. Superconducting qubits (such as charge, flux,
and transmon qubits) behave as artificial atoms, they have relatively long
decoherence times [3-7], and various single- and multiple-qubit operations
with state readout have been demonstrated [8-12]. On the other hand, a
superconducting resonator provides a quantized cavity field which acts as a
quantum bus and thus can mediate long-distance and strong interaction
between distant superconducting qubits [13-15]. Furthermore, the strong
coupling between a microwave cavity and superconducting charge qubits [16]
or flux qubits [17] was earlier predicated in theory and has been
experimentally demonstrated [18,19]. Because of these features, circuit QED
has been widely utilized for quantum information processing. During the past
decade, based on circuit QED, many theoretical proposals have been presented
for the preparation of Fock states, coherent states, squeezed states,
Sch\"ordinger Cat states, and arbitrary superpositions of Fock states of a
single superconducting resonator [20-22]. So far, Fock states and their
superpositions of a resonator have been experimentally produced by using a
superconducting qubit [23-25].

Intense effort has been recently devoted to the preparation of entangled
states of photons in two or more superconducting resonators [26-29]. The
NOON states are a special type of photonic entangled states with two
orthogonal component states in maximal superposition, which play the crucial
role in quantum optical lithography [30,31], quantum metrology [32-35],
precision measurement of transmons [36-38], and quantum information
processing [39,40].

In Ref. [26], a theoretical method for synthesizing an arbitrary quantum
state of two superconducting resonators using a tunable superconducting
qubit has been proposed. This method is based on alternative resonant
interactions of the coupler qubit with two cavity modes and a classical
pulse. As pointed out in [26], the Rabi frequency of the classical pulse
needs to be much smaller than the photon-number-dependent Stark shifts
induced by dispersive interaction with the two field modes and, hence, the
pulse can drive the qubit to undergo a rotation conditional upon the state
of the cavity modes. This implies that the time needed to complete the
rotation in each step should be much (two orders of magnitude) longer than
the vacuum Rabi period of the coupled qubit-resonator system.

In Ref. [27], the authors proposed a theoretical scheme for creating NOON
states of two resonators, which was implemented in experiments for $N\leq 3$
by H. Wang et al. [28]. The method in [27,28] operates essentially by
employing two three-level superconducting qutrits as couplers, preparing
them in a Bell state, and then performing $N$ steps of operation to swap the
coherence of the Bell state onto the two resonators through a sequence of
classical pulses applied to the two coupler qutrits. In addition, as
discussed in [27,28], a third resonator or cavity is needed in order to
prepare the two coupler qutrits in the Bell state.

Ref. [29] presented an approach to control the quantum state of two
superconducting resonators using a complicated classical microwave pulse.
For the generation of NOON states, this scheme also requires two
superconducting qubits which are initially prepared in a Bell state. Another
problem is that the produced state is essentially an entangled state of two
resonators and two qubits. To obtain the pure photonic NOON state, one
should use additional techniques to decouple the qubits from the resonators.

As reported in [28], the fidelity of the obtained NOON state decreases
dramatically with the photon number $N$ due to decoherence, dropping to 0.33
for $N=3$. In order to be useful in quantum technologies, the fidelity needs
to be significantly improved. Thus, it is worthy of exploring more efficient
schemes to generate the NOON states with a higher fidelity.

In this work, we propose an alternative scheme for generating the NOON state
of two resonators coupled to a superconducting transmon qutrit, via resonant
interactions. This proposal has the following advantages: (i) Because of
using only one superconducting qutrit and two resonators, the experimental
setup is greatly simplified when compared with that in [27-29], which is
important for decreasing decoherence effects; (ii) In principle, there is no
limitation on the intensity of the classical pulse for our scheme and thus
the operation can be performed much faster when compared with the method in
[26]. Overall, the important features of our scheme are simplicity,
rapidness, and robustness.

We will also give a detailed discussion of the experimental issues and then
analyze the possible experimental implementation. Our numerical simulation
shows that a high-fidelity generation of the NOON state with $N\leq 3$ is
feasible within the present circuit QED technique.

\section*{Results}

\noindent \textbf{Noon-state preparation.} Consider two resonators coupled
to a superconducting transmon qutrit (Fig. 1). The three ladder-type levels
of the qutrit are labeled as $\left| g\right\rangle ,\left| e\right\rangle ,$%
and $\left| f\right\rangle $ with energy $E_g<E_e<E_f.$ Suppose that the
coupler qutrit is initially in the state $\frac 1{\sqrt{2}}(\left|
f\right\rangle +\left| e\right\rangle )$ and the two resonators are
initially in the vacuum state $\left| 0\right\rangle _a\left| 0\right\rangle
_b$. The qutrit can be made to be decoupled from the two resonators by a
prior adjustment of the qutrit level spacings. Note that for superconducting
transmon qutrits, the level spacings can be rapidly adjusted by varying
external control parameters (e.g., magnetic flux applied to superconducting
quantum interference device (SQUID) loops of two-junction transmon qutrits;
see, e.g., [41-43]).

For simplicity, we define $\omega _{eg}$ ($\omega _{fe}$) as the $\left|
g\right\rangle \leftrightarrow \left| e\right\rangle $ ($\left|
e\right\rangle \leftrightarrow \left| f\right\rangle $) transition frequency
of the qutrit and $\Omega _{eg}$ ($\Omega _{fe}$) as the Rabi frequency of
the classical pulse driving the coherent $\left| g\right\rangle
\leftrightarrow \left| e\right\rangle $ ($\left| e\right\rangle
\leftrightarrow \left| f\right\rangle $) transition. In addition, the
frequency, initial phase, and duration of the microwave pulse are denoted as
\{$\omega ,$ $\varphi ,$ $t$\} in the rest of the paper.

The procedure for generating the NOON state of photons in the two resonators
contains $2N$ steps. We assume that resonator $b$ ($a$) is decoupled from
the qutrit during each of the first (second) $N$ steps\ due to large
detunings, which can be achieved by prior adjustment of the resonator
frequency. The effects of off-resonant qutrit-resonator couplings and
classical drivings on the fidelity of the prepared state will be taken into
account later.

Before the operations for the first $N$ steps, we need to adjust the level
spacings of the qutrit such that resonator $a$ is resonant with the $\left|
g\right\rangle \leftrightarrow \left| e\right\rangle $\ transition, but is
far off-resonant with (decoupled from)\ the $\left| e\right\rangle
\leftrightarrow \left| f\right\rangle $\ transition so that the coupling
between resonator $a$\ and the $\left| e\right\rangle \leftrightarrow \left|
f\right\rangle $\ transition\ can be neglected [Fig.~2(a)]. Meanwhile,
resonator $b$\ is far off-resonant with both of these two transitions and
thus it is unaffected during this interaction (i.e., resonator $b$ is
decoupled from the qutrit). Under these conditions, the state $\left|
f\right\rangle $ remains unchanged due to the large detuning. In the
interaction picture with respect to the free Hamiltonian of the whole
system, the Hamiltonian describing this operation is given by $H_I=\hbar
\left( g_{eg}a^{+}\left| g\right\rangle \left\langle e\right| \right) +h.c.,$
where $a^{+}$ is the photon creation operator of the mode of resonator $a$,
and $g_{eg}$ is the coupling constant between the mode of the resonator $a$
and the $\left| g\right\rangle \leftrightarrow \left| e\right\rangle $
transition [Fig.~2(a)].

The operations of the first $N$ steps are described below:

Step $1$: Let resonator $a$ resonant with the $\left| g\right\rangle
\leftrightarrow \left| e\right\rangle $\ transition. Under the Hamiltonian $%
H_I$, the state component $\left| f\right\rangle \left| 0\right\rangle _a$
is not changed because of $H_I\left| f\right\rangle \left| 0\right\rangle
_a=0$, while $\left| e\right\rangle \left| 0\right\rangle _a$ undergoes the
Jaynes-Cumming evolution [44]. After an interaction time $t_1=\pi /(2g_{eg})$
(i.e., half a Rabi oscillation), the state $\left| e\right\rangle \left|
0\right\rangle _a$ changes to $-i\left| g\right\rangle \left| 1\right\rangle
_a$ (for the details, see the discussion in the part of Methods below).
Hence, the initial state $\frac 1{\sqrt{2}}(\left| f\right\rangle +\left|
e\right\rangle )\left| 0\right\rangle _a\left| 0\right\rangle _b$ of the
whole system becomes \textbf{\ }

\begin{equation}
\frac 1{\sqrt{2}}(\left| f\right\rangle \left| 0\right\rangle _a-i\left|
g\right\rangle \left| 1\right\rangle _a)\left| 0\right\rangle _b.
\end{equation}
Then, apply a microwave pulse of $\{\omega _{eg},$ $-\pi /2,$ $\pi /\left(
2\Omega _{eg}\right) \}$ to the qutrit to pump the state $\left|
g\right\rangle $ back to $\left| e\right\rangle $ [Fig.~2(b)], transforming
the state (1) to
\begin{equation}
\frac 1{\sqrt{2}}(\left| f\right\rangle \left| 0\right\rangle _a-i\left|
e\right\rangle \left| 1\right\rangle _a)\left| 0\right\rangle _b.
\end{equation}
Here and below, we assume $\Omega _{eg}\gg g_{eg}$ so that the interaction
between the qutrit and resonotor\textbf{\ }$a$ is negligible during the
application of this pulse.

Step $j$ ($j=2,3,...,N-1$):\textbf{\ }Repeat the operation of step 1. The
time for the qutrit interacting with resonator $a$ is set by $t_{j}=\pi
/\left( 2\sqrt{j}g_{eg}\right) $ (i.e., half a Rabi oscillation). After an
interaction time $t_{j},$ the state $\left\vert f\right\rangle \left\vert
0\right\rangle _{a}$ remains unchanged while the state $\left\vert
e\right\rangle \left\vert j-1\right\rangle _{a}$ changes to $-i\left\vert
g\right\rangle \left\vert j\right\rangle _{a},$ which further changes to $%
-i\left\vert e\right\rangle \left\vert j\right\rangle _{a}$ due to a
microwave pulse of $\{\omega _{eg},$ $-\pi /2,$ $\pi /\left( 2\Omega
_{eg}\right) \}$ pumping the state $\left\vert g\right\rangle $ back to $%
\left\vert e\right\rangle $. Hence, one can easily verify that after the
operation of steps ($2,3,...,N-1$), the state (2) becomes
\begin{equation}
\frac{1}{\sqrt{2}}(\left\vert f\right\rangle \left\vert 0\right\rangle
_{a}+(-i)^{N-1}\left\vert e\right\rangle \left\vert N-1\right\rangle
_{a})\left\vert 0\right\rangle _{b}.
\end{equation}

Step $N$: Let resonator $a$\textbf{\ }resonant with the $\left|
g\right\rangle \leftrightarrow \left| e\right\rangle $\ transition for an
interaction time $t_N=\pi /\left( 2\sqrt{N}g_{eg}\right) $ [Fig.~2(a)]. As a
result, we have the transformation $\left| e\right\rangle \left|
N-1\right\rangle _a\rightarrow $ $-i\left| g\right\rangle \left|
N\right\rangle _a$ while the state $\left| f\right\rangle \left|
0\right\rangle _a$ remains unchanged. Thus, the state (3) becomes
\begin{equation}
\frac 1{\sqrt{2}}(\left| f\right\rangle \left| 0\right\rangle
_a+(-i)^N\left| g\right\rangle \left| N\right\rangle _a)\left|
0\right\rangle _b.
\end{equation}

In above we have given a detailed description of the operations for the
first $N$ steps. Now let us give a description on the second $N$ steps. To
begin with, we need to adjust the level spacings of the qutrit to bring
resonator $b$\ resonant with the $\left| e\right\rangle \leftrightarrow
\left| f\right\rangle $\ transition but far off-resonant with the $\left|
g\right\rangle \leftrightarrow \left| e\right\rangle $\ transition
[Fig.~2(c)]. On the other hand, resonator $a$\ is far off-resonant with each
transition so that it is unaffected during this interaction (i.e., resonator
$a$ is decoupled from the qutrit, which can be achieved by adjusting the
frequency of resonator $a$). In the interaction picture with respect to the
free Hamiltonian of the whole system, the Hamiltonian governing this
operation is given by $H_I=\hbar \left( g_{fe}b^{+}\left| e\right\rangle
\left\langle f\right| \right) +h.c.$, where $b^{+}$\ is the photon creation
operator of the mode of resonator $b$, and $g_{fe}$\ is the coupling
constant between the resonator $b$\ and the $\left| e\right\rangle
\leftrightarrow \left| f\right\rangle $\ transition [Fig.~2(c)].

Since the level spacings of the qutrit are now different from those used in
the operation of the first $N$ steps, we now define $\omega _{eg}^{\prime},
\omega _{fe}^{\prime}$, and $\omega_{fg}^{\prime}$ as the $%
\left|g\right\rangle \leftrightarrow \left| e\right\rangle $ transition
frequency, the $\left|e\right\rangle \leftrightarrow \left| f\right\rangle $
transition frequency, and the $\left|g\right\rangle \leftrightarrow \left|
f\right\rangle $ transition frequency of the qutrit, respectively.

The operations of the second $N$ steps are as follows:

Step $1$: Let resonator $b$\ resonant with the $\left| e\right\rangle
\leftrightarrow \left| f\right\rangle $\ transition [Fig.~2(c)]. Under the
Hamiltonian $H_I$, the state component $\left| g\right\rangle \left|
0\right\rangle _b$ does not change because of $H_I\left| g\right\rangle
\left| 0\right\rangle _b=0$, while $\left| f\right\rangle \left|
0\right\rangle _b$ undergoes the Jaynes-Cumming evolution. After an
interaction time $\mathbf{t}_1^{\prime }\mathbf{=\pi /(2g}_{fe}\mathbf{)}$,
the state $\left| f\right\rangle \left| 0\right\rangle _b$ changes to $%
-i\left| e\right\rangle \left| 1\right\rangle _b$ (see the discussion in the
Methods below). Thus, one can see that the state (4) changes to \textbf{\ }
\begin{equation}
\frac 1{\sqrt{2}}[-i\left| e\right\rangle \left| 0\right\rangle _a\left|
1\right\rangle _b+(-i)^N\left| g\right\rangle \left| N\right\rangle _a\left|
0\right\rangle _b].
\end{equation}
Then, apply a microwave pulse of $\{\omega _{fe}^{\prime },$\ $-\pi /2,$\ $%
\pi /\left( 2\Omega _{fe}\right) \}$\ to the qutrit to pump the state $%
\left| e\right\rangle $\ back to $\left| f\right\rangle $\ [Fig.~2(d)],
transforming the state (5) to
\begin{equation}
\frac 1{\sqrt{2}}[-i\left| f\right\rangle \left| 0\right\rangle _a\left|
1\right\rangle _b+(-i)^N\left| g\right\rangle \left| N\right\rangle _a\left|
0\right\rangle _b].
\end{equation}
Here and below, we assume $\Omega _{fe}\gg g_{fe}$ such that\textbf{\ }the
qutrit-resonator coupling is negligible during the application of this pulse.

Step $j$\ ($j=2,3,...,N-1$): Repeat the operation of step $1$. The time for
the qutrit interacting with resonator $b$ is set by $t_j^{\prime }=\pi
/\left( 2\sqrt{j}g_{fe}\right) $. After an interaction time $t_j^{\prime },$
the state $\left| g\right\rangle \left| 0\right\rangle _b$ remains unchanged
while the state $\left| f\right\rangle \left| j-1\right\rangle _b$ changes
to $-i\left| e\right\rangle \left| j\right\rangle _b,$ which further turns
into $-i\left| f\right\rangle \left| j\right\rangle _b$ because of a
microwave pulse of $\{\omega _{fe}^{\prime },$ $-\pi /2,$ $\pi /\left(
2\Omega _{fe}\right) \}$ pumping the state $\left| e\right\rangle $ back to $%
\left| f\right\rangle $. After step\textbf{\ }$N-1$\textbf{, }the state (6)
becomes

\begin{equation}
\frac{1}{\sqrt{2}}[\left( -i\right) ^{N-1}\left\vert f\right\rangle
\left\vert 0\right\rangle _{a}\left\vert N-1\right\rangle
_{b}+(-i)^{N}\left\vert g\right\rangle \left\vert N\right\rangle
_{a}\left\vert 0\right\rangle _{b}].
\end{equation}

Step $N$: Apply a microwave pulse of $\{\omega _{eg}^{\prime },$ $-\pi /2,$ $%
\pi /\left( 2\Omega _{eg}\right) \}$ to the qutrit to pump the state $\left|
g\right\rangle $ back to $\left| e\right\rangle $ [note that in Fig.~2(d),
the pulse is now resonant to the $\left| g\right\rangle \leftrightarrow
\left| e\right\rangle $\ transition, instead of the $\left| e\right\rangle
\leftrightarrow \left| f\right\rangle $\ transition]$.$ To neglect the%
\textbf{\ }qutrit-resonator coupling\textbf{\ }during this pulse, the
condition $\Omega _{eg}\gg g_{fe}$ needs to be satisfied. Then, let
resonator $b$\ resonant with the $\left| e\right\rangle \leftrightarrow
\left| f\right\rangle $\ transition\textbf{\ }for an interaction time $%
t_N^{\prime }=\pi /\left( 2\sqrt{N}g_{fe}\right) $ [Fig.~2(c)], leading to
the transformation $\left| f\right\rangle \left| N-1\right\rangle
_b\rightarrow -i\left| e\right\rangle \left| N\right\rangle _b$. Meanwhile,
resonator $a$ remains decoupled from the qutrit. As a result, the state (7)
changes to
\begin{equation}
\frac 1{\sqrt{2}}(-i)^N[\left| 0\right\rangle _a\left| N\right\rangle
_b+\left| N\right\rangle _a\left| 0\right\rangle _b]\left| e\right\rangle .
\end{equation}
Then, adjust the level spacings of the qutrit back to the original level
configuration such that the qutrit is decoupled from the two resonators. The
result (8) shows that the two resonators $a$\ and $b$\ are prepared in a
NOON state of photons, which are disentangled from the qutrit.

Previously we have assumed that during the first (second) $N$ steps of
operations, the resonator $b$ ($a$) is decoupled from the qutrit. In
principle, this requirement can be met by adjusting the level spacings of
the qutrit [41-43] or the resonator mode frequency such that the irrelevant
resonator during the operation is highly detuned from the transition between
any two levels of the coupler qutrit. The rapid tuning of cavity frequencies
has been demonstrated in superconducting microwave cavities (e.g., in less
than a few nanoseconds for a superconducting transmission line resonator
[45]).

As shown above, our NOON-state preparation is based on the following
approximations. For the first $N$ steps of operation, we have neglected the
off-resonant interaction between resonator $a$ and the $\left|
e\right\rangle \leftrightarrow \left| f\right\rangle $ transition of the
qutrit, the off-resonant interaction between the pulse and the $\left|
e\right\rangle \leftrightarrow \left| f\right\rangle $ transition of the
qutrit, and the off-resonant coupling of resonator $b$ with the transition
between any two levels of the qutrit. For the second $N$ steps of operation,
we have omitted the off-resonant interaction between resonator $b$ and the $%
\left| g\right\rangle \leftrightarrow \left| e\right\rangle $ transition of
the qutrit, the off-resonant interaction between the pulse and the $\left|
g\right\rangle \leftrightarrow \left| e\right\rangle $ transition of the
qutrit, and the off-resonant coupling of resonator $a$ with the transition
between any two levels of the qutrit. In addition, for each step of
operation, there exists an inter-cavity cross coupling, which was also not
considered in our NOON-state preparation above. To quantify how well our
protocol works out, later we will perform a numerical simulation for $N\leq
5,$ by taking all these effects into account.

\noindent \textbf{Experimental issues.} For the method to work the primary
considerations shall be given to:

(i) The total operation time $\tau ,$ given by
\begin{equation}
\tau =\sum_{j=1}^N\pi /(2\sqrt{j}g_{eg})+\sum_{j=1}^N\pi /\left( 2\sqrt{j}%
g_{fe}\right) +N\pi /\left( 2\Omega _{eg}\right) +\left( N-1\right) \pi
/\left( 2\Omega _{fe}\right) +3t_d
\end{equation}
(where $t_d\sim 1-3$ ns is the typical time required for adjusting the
qutrit level spacings), needs to be much shorter than the energy relaxation
time $T_1$ ($T_1^{^{\prime }}$) and dephasing time $T_2$ ($T_2^{^{\prime }}$%
) of the level $\left| f\right\rangle $ ($|e\rangle $) of the qutrit$,$ such
that decoherence caused by energy relaxation and dephasing of the qutrit is
negligible for the operation. Note that $T_1^{^{\prime }}$ and $T_2^{\prime
} $ of the qutrit are comparable to $T_1$ and $T_2,$ respectively. For
instance, $T_1^{\prime }\sim 2T_1$ and $T_2^{^{\prime }}\sim T_2$ for
transmon qutrits.

(ii) For resonator $k$\ ($k=a,b$), the lifetime of the resonator mode is
given by $T_{cav}^k=\left( Q_k/2\pi \nu _k\right) /\overline{n}_k,$\ where $%
Q_k$, $\nu _k$\ and $\overline{n}_k$\ are the (loaded) quality factor,
frequency, and the average photon number of resonator $k$, respectively. For
the two resonators, the lifetime of entanglement of the resonator modes is
given by
\begin{equation}
T_{cav}\mathbf{=}\frac 12\min \mathbf{\{}T_{cav}^a\mathbf{,}T_{cav}^b\mathbf{%
\},}
\end{equation}
which should be much longer than $\tau ,$ such that the effect of resonator
decay is negligible during the operation.

(iii) The inter-cavity cross coupling between the two resonators is
determined mostly by the coupling capacitance $C_c$ and the qutrit's self
capacitance $C_q$, because the\textbf{\ }field leakage through space\textbf{%
\ }is extremely low for high-$Q$\ resonators as long as the inter-cavity
distance is much greater than transverse dimension of the cavities - a
condition easily met in experiments for the two resonators. Furthermore, as
the result of our numerical simulation shown below (see Fig. 4), the effects
of the inter-cavity coupling can however be made negligible as long as the
corresponding inter-cavity coupling constant $g_{ab}$ between resonators%
\textrm{\ }$a$ and $b$ is sufficiently small.

\noindent \textbf{Fidelity.} Hereafter, we give a discussion of the fidelity
of the prepared NOON state for $N\leq 5.$

The first $N$ steps above for creating the NOON state involves the following
two basic types of interactions:

(i) The first one is the resonant coupling between resonator $a$\ and the $%
\left\vert g\right\rangle \leftrightarrow \left\vert e\right\rangle $\
transition. When the interaction between resonator $a$ and the $\left\vert
e\right\rangle \leftrightarrow \left\vert f\right\rangle $\ transition [Fig.
3(a)], the coupling between resonator $b$ and the qutrit [Fig. 3(a)], and
the inter-cavity crosstalk between the two resonators are taken into
account, the corresponding interaction Hamiltonian is thus given by
\begin{eqnarray}
H_{I,1} &=&\hbar \left( g_{eg}a\left\vert e\right\rangle \left\langle
g\right\vert +h.c.\right) +\hbar \left( \widetilde{g}_{fe}e^{i\delta
_{1}t}a\left\vert f\right\rangle \left\langle e\right\vert +h.c.\right)
\nonumber \\
&&\ \ +\hbar \left( \mu _{eg}e^{i\delta _{eg}t}b\left\vert e\right\rangle
\left\langle g\right\vert +h.c.\right) +\hbar \left( \mu _{fe}e^{i\delta
_{fe}t}b\left\vert f\right\rangle \left\langle e\right\vert +h.c.\right)
\nonumber \\
&&\ \ +\hbar g_{ab}\left( e^{i\Delta t}ab^{+}+h.c.\right) .
\end{eqnarray}%
where the first term represents the resonant interaction of resonator $a$
with the $\left\vert g\right\rangle \leftrightarrow \left\vert
e\right\rangle $ transition, while the second term represents the unwanted
off-resonant coupling between resonator $a$ and the $\left\vert
e\right\rangle \leftrightarrow \left\vert f\right\rangle $ transition with
coupling constant $\widetilde{g}_{fe}$ and detuning\textbf{\ }$\delta
_{1}=\omega _{fe}-\omega _{a}<0$ [Fig.~3(a)]. In addition, the third term
represents the unwanted off-resonant coupling between resonator $b$ and the $%
\left\vert g\right\rangle \leftrightarrow \left\vert e\right\rangle $
transition with coupling constant $\mu _{eg}$ and detuning\textbf{\ }$\delta
_{eg}=\omega _{eg}-\omega _{b}>0$ [Fig.~3(a)], while the fourth term
represents the unwanted off-resonant coupling between resonator $b$ and the $%
\left\vert e\right\rangle \leftrightarrow \left\vert f\right\rangle $
transition with coupling constant $\mu _{fe}$ and detuning\textbf{\ }$\delta
_{fe}=\omega _{fe}-\omega _{b}>0$ [Fig.~3(a)]. Finally, the last term
indicates the inter-cavity crosstalk between the two resonators, where $%
\Delta =\omega _{b}-\omega _{a}<0$ is the detuning between the two
resonators. The Hamiltonian\textbf{\ }$H_{I,1}$\textbf{\ }here, together with%
\textbf{\ }$H_{I,2}$\textbf{, }$H_{I,3}$\textbf{\ }and\textbf{\ }$H_{I,4}$%
\textbf{\ }below, is written in the interaction picture with respect to the
free Hamiltonian of the whole system.

(ii) The second one corresponds to the application of the pulse with $%
\{\omega _{eg},$\ $-\pi /2,$\ $\pi /\left( 2\Omega _{eg}\right) \}$\ to the
qutrit. The interaction Hamiltonian governing this basic operation is given
by
\begin{eqnarray}
H_{I,2} &=&\hbar \left( \Omega _{eg}e^{-i\pi /2}\left\vert g\right\rangle
\left\langle e\right\vert +h.c.\right) +\hbar \left[ \widetilde{\Omega }%
_{fe}e^{i\left( -\delta _{2}t-\pi /2\right) }\left\vert e\right\rangle
\left\langle f\right\vert +h.c.\right]   \nonumber \\
&&\ +H_{I,1},
\end{eqnarray}%
where the first term represents the resonant interaction of the pulse with
the $\left\vert g\right\rangle \leftrightarrow \left\vert e\right\rangle $
transition, while the second one represents the unwanted off-resonant
coupling between the pulse and the $\left\vert e\right\rangle
\leftrightarrow \left\vert f\right\rangle $ transition with Rabi frequency $%
\widetilde{\Omega }_{fe}$ and detuning\textbf{\ }$\delta _{2}=\omega
_{fe}-\omega _{eg}<0$ [Fig.~3(b)]. Here, $H_{I,1}$\ is the Hamiltonian given
in Eq. (11), describing the coupling between resonator $a$ and the qutrit,
the coupling between resonator $b$ and the qutrit, as well as the
inter-cavity crosstalk between the two resonators during the pulse.

The second $N$ steps above for creating the NOON state covers the following
two basic types of interactions:

(iii) The third one corresponds to the resonant coupling between the
resonator $b$\ and the $\left\vert e\right\rangle \leftrightarrow \left\vert
f\right\rangle $\ transition. When the unwanted off-resonant coupling
between this resonator and the $\left\vert g\right\rangle \leftrightarrow
\left\vert e\right\rangle $\ transition [Fig. 3(c)], the coupling between
resonator $a$ and the qutrit [Fig. 3(c)], and the inter-cavity crosstalk
between the two resonators are considered, the total interaction Hamiltonian
reads
\begin{eqnarray}
H_{I,3} &=&\hbar \left( g_{fe}b\left\vert f\right\rangle \left\langle
e\right\vert +h.c.\right) +\hbar \left( \widetilde{g}_{eg}e^{i\delta
_{3}t}b\left\vert e\right\rangle \left\langle g\right\vert +h.c.\right)
\nonumber \\
&&\ +\hbar \left( \widetilde{\mu }_{eg}e^{i\widetilde{\delta }%
_{eg}t}a\left\vert e\right\rangle \left\langle g\right\vert +h.c.\right)
+\hbar \left( \widetilde{\mu }_{fe}e^{i\widetilde{\delta }_{fe}t}a\left\vert
f\right\rangle \left\langle e\right\vert +h.c.\right)   \nonumber \\
&&\ +\hbar g_{ab}\left( e^{i\Delta t}ab^{+}+h.c.\right) ,
\end{eqnarray}%
where the first term represents the resonant interaction of resonator $b$
with the $\left\vert e\right\rangle \leftrightarrow \left\vert
f\right\rangle $ transition, while the second term represents the unwanted
off-resonant coupling between resonator $b$ and the $\left\vert
g\right\rangle \leftrightarrow \left\vert e\right\rangle $ transition with
coupling constant $\widetilde{g}_{eg}$ and detuning\textbf{\ }$\delta
_{3}=\omega _{eg}^{\prime }-\omega _{b}>0$ [Fig.~3(c)]. In addition, the
third term represents the unwanted off-resonant coupling between resonator $a
$ and the $\left\vert g\right\rangle \leftrightarrow \left\vert
e\right\rangle $ transition with coupling constant $\widetilde{\mu }_{eg}$
and detuning\textbf{\ }$\widetilde{\delta }_{eg}=\omega _{eg}^{\prime
}-\omega _{a}<0$ [Fig.~3(c)], while the fourth term represents the unwanted
off-resonant coupling between resonator $a$ and the $\left\vert
e\right\rangle \leftrightarrow \left\vert f\right\rangle $ transition with
coupling constant\textbf{\ }$\widetilde{\mu }_{fe}$\textbf{\ }and detuning%
\textbf{\ }$\widetilde{\delta }_{fe}=\omega _{fe}^{\prime }-\omega _{a}<0$%
\textbf{\ }[Fig.~3(c)]\textbf{.}

(iv) The last one is the pump of the qutrit with the pulse $\{\omega
_{fe}^{\prime },$\ $-\pi /2,$\ $\pi /\left( 2\Omega _{fe}\right) \}$, with
the interaction Hamiltonian described by
\begin{eqnarray}
H_{I,4} &=&\hbar \left( \Omega _{fe}e^{-i\pi /2}\left\vert e\right\rangle
\left\langle f\right\vert +h.c.\right) +\hbar \left[ \widetilde{\Omega }%
_{eg}e^{i\left( -\delta _{4}t-\pi /2\right) }\left\vert g\right\rangle
\left\langle e\right\vert +h.c.\right]   \nonumber \\
&&+H_{I,3},
\end{eqnarray}%
where the first term denotes the resonant pump of the $\left\vert
e\right\rangle \leftrightarrow \left\vert f\right\rangle $\ transition,
while the second one represents the unwanted off-resonant excitation of the $%
\left\vert g\right\rangle \leftrightarrow \left\vert e\right\rangle $\
transition with Rabi frequency $\widetilde{\Omega }_{eg}$\ and detuning $%
\delta _{4}=\omega _{eg}^{\prime }-\omega _{fe}^{\prime }>0$\ [Fig.~3(d)].
Here, $H_{I,3}$\ is the Hamiltonian given in Eq. (13), describing the
coupling between resonator $a$ and the qutrit, the coupling between
resonator $b$ and the qutrit, as well as the inter-cavity crosstalk between
the two resonators during the pulse.

It is noted that the term describing the pulse- or resonator-induced
coherent $\left| g\right\rangle \leftrightarrow \left| f\right\rangle $\
transition for the qutrit is not included in the Hamiltonians $H_{I,1},$\ $%
H_{I,2}$\ $H_{I,3},$\ and $H_{I,4},$ since the error caused by this
transition is much smaller than those described above. This is because: (i)
the two resonators and the pulses are highly detuned from the $\left|
g\right\rangle \leftrightarrow \left| f\right\rangle $\ transition due to $%
\omega _a,\omega _b,\omega \ll \omega _{fg},$ $\omega _{fg}^{\prime }$
(Fig.~3); and (ii) for a transmon qutrit with the three levels considered
here, the $\left| g\right\rangle \leftrightarrow \left| f\right\rangle $
dipole matrix element is much smaller than that of the $\left|
g\right\rangle \leftrightarrow \left| e\right\rangle $ and $\left|
e\right\rangle \leftrightarrow \left| f\right\rangle $ transitions [46].

When the dissipation and dephasing are included, the dynamics for the
\textit{k}th type of interactions is determined by the following master
equation
\begin{eqnarray}
\frac{d\rho }{dt} &=&-i\left[ H_{I,k},\rho \right] +\kappa _a\mathcal{L}%
\left[ a\right] +\kappa _b\mathcal{L}\left[ b\right] +  \nonumber \\
&&+\gamma _{fe}\mathcal{L}\left[ S_{-,fe}\right] +\gamma _{eg}\mathcal{L}%
\left[ S_{-,eg}\right]  \nonumber \\
&&+\gamma _{\varphi ,f}\left( S_{ff}\rho S_{ff}-S_{ff}\rho /2-\rho
S_{ff}/2\right)  \nonumber \\
&&+\gamma _{\varphi ,e}\left( S_{ee}\rho S_{ee}-S_{ee}\rho /2-\rho
S_{ee}/2\right) ,
\end{eqnarray}
where $H_{I,k}$ for $k=1, 2, 3$, and $4$ are the above Hamiltonians $%
H_{I,1}, H_{I,2}, H_{I,3}$, and $H_{I,4}$, respectively; $\mathcal{L}\left[
\Lambda \right] =\Lambda \rho \Lambda ^{+}-\Lambda ^{+}\Lambda \rho /2-\rho
\Lambda ^{+}\Lambda /2$ (with $\Lambda =a,b,S_{-,fe},S_{-,fg},S_{-,eg}$),\ $%
S_{ff}=\left| f\right\rangle \left\langle f\right| $, and $S_{ee}=\left|
e\right\rangle \left\langle e\right| $. In addition, $\kappa _a$\ ($\kappa
_b $) is the decay rate of the resonator mode $a$\ ($b$);\ $\gamma _{fe}$\
is the energy relaxation rate for the level $\left| f\right\rangle $\
associated with the decay path $\left| f\right\rangle \rightarrow \left|
e\right\rangle $; $\gamma _{eg}$\ is that for the level $\left|
e\right\rangle $; and $\gamma _{\varphi ,f}$ ($\gamma _{\varphi ,e}$) is the
dephasing rate of the level $\left| f\right\rangle $ ($\left| e\right\rangle
$)\textbf{.}

The fidelity of the whole operation is given by $\mathcal{F}=\left\langle
\psi _{id}\right| \widetilde{\rho }\left| \psi _{id}\right\rangle ,$ where $%
\left| \psi _{id}\right\rangle $ is the output state given in Eq.~(8) for an
ideal system (i.e., without unwanted couplings\textbf{,} dissipation, and
dephasing) after the entire operation, while $\widetilde{\rho }$ is the
final density operator of the whole system when the operations are performed
in a realistic physical system.

We now numerically calculate the fidelity of the NOON state prepared above,
with $N\leq 5$. For simplicity, we set: (i) $\delta _1/\left( 2\pi \right)
=\delta _2/\left( 2\pi \right) =-400$ MHz, and $\delta _3/\left( 2\pi
\right) =\delta _4/\left( 2\pi \right) =400$ MHz $[43],$ (ii) $\Omega
_{eg}=\Omega _{fe}=\Omega $ (achievable via adjusting the pulse intensities)$%
,$ thus $\widetilde{\Omega }_{fe}\sim \sqrt{2}\Omega ,$ and $\widetilde{%
\Omega }_{eg}\sim \Omega /\sqrt{2}$ for the transmon qutrit here [46]; (iii)
$g_{eg}=\widetilde{g}_{eg}=g,$ and thus $g_{fe}\sim \widetilde{g}_{fe}\sim
\sqrt{2}g$ [46]. In addition, $\mu _{eg}$, $\mu _{fe},$ $\widetilde{\mu }%
_{eg},$ and $\widetilde{\mu }_{eg}$ can be determined due to $\mu _{eg}\sim
g_{eg}\sqrt{\omega _b/\omega _a},$ $\mu _{fe}\sim \widetilde{g}_{fe}\sqrt{%
\omega _b/\omega _a},$ $\widetilde{\mu }_{eg}\sim \widetilde{g}_{eg}\sqrt{%
\omega _a/\omega _b},$ and $\widetilde{\mu }_{fe}\sim g_{fe}\sqrt{\omega
_a/\omega _b}$. For superconducting transmon qutrits, the typical transition
frequency between two neighbor levels is between 5 and 10 GHz. As an
example, let us consider resonator $a$ with frequency $\omega _a/\left( 2\pi
\right) \sim 6$ GHz while resonator $b$ with frequency $\omega _b/\left(
2\pi \right) \sim 3.5$ GHz. Other parameters used in the numerical
calculation are as follows: (i) $\Delta /\left( 2\pi \right) =-2.5$ GHz, (ii)%
$\ \Omega /\left( 2\pi \right) =18$ MHz [47,48]), (iii) $\gamma _{\varphi
,f}^{-1}=\gamma _{\varphi ,e}^{-1}=3$ $\mu $s, $\gamma _{fe}^{-1}=1.5$ $\mu $%
s, $\gamma _{eg}^{-1}=3$ $\mu $s (which are available in experiment [42]),
and (iv) $\kappa _a^{-1}=\kappa _b^{-1}=20$ $\mu $s. For the parameters
chosen here, the fidelity for $N\leq 5$ is shown in Fig.~4 for $g_{ab}=0,$ $%
g,$ and $2g.$ Fig. 4 was plotted by numerically optimizing the coupling
constants, e.g., $g/\left( 2\pi \right) =3.9$, $2.2$, $1.8$, $1.5$, $1.3$
MHz for $N=1,2,3,4,5,$ respectively. The coupling strengths with these
values are readily achievable in experiment because $g/\left( 2\pi \right)
\sim 220$\textbf{\ }MHz has been reported for a superconducting transmon
qubit coupled to a one-dimensional standing-wave CPW (coplanar waveguide)
resonator [49].\textbf{\ }It can be seen from Fig. 4 that when $g_{ab}\leq
2g,$ the effect of the inter-cavity coupling is negligible and a high
fidelity $\gtrsim $ $76\%$ can be obtained for $N\leq 3.$

The fidelity can be further increased by improving the system parameters.
For instance, Fig. 5 shows that the fidelity for $N=3$ can be increased to $%
\sim 85\%$ for $\gamma _{\varphi ,f}^{-1}=\gamma _{\varphi ,e}^{-1}=T=10$ $%
\mu $s, $\gamma _{fe}^{-1}=T/2$, and $\gamma _{eg}^{-1}=T$, which can be
reached in the near future due to the rapid development of the circuit-QED
techniques (e.g., decoherence time $\sim 10$ $\mu $s has been demonstrated
in a superconducting transmon qubit coupled to a 3D cavity [50]).

For the resonators $a$ and $b$ of frequencies given above and the $\kappa
_a^{-1}$ and $\kappa _b^{-1}$ used in the numerical calculation, the
required quality factors for the two resonators are $Q_a\sim 7.5\times 10^5$
and $Q_b\sim 4.4\times 10^5$. Note that superconducting CPW resonators with
a loaded quality factor $Q\sim 10^6$ have been experimentally demonstrated
[51,52], and planar superconducting resonators with internal quality factors
above one million ($Q>10^6$) have also been reported recently [53]. Our
analysis given here demonstrates that high-fidelity generation of the NOON
state with $N\leq 3$ using the present proposal is possible within the
present circuit QED techniques.

The condition, $g_{ab}\leq 2g,$ is not difficult to satisfy with typical
capacitive cavity-qutrit coupling illustrated in Fig. 1. As discussed in
[54], as long as the cavities are physically well separated, the
inter-cavity crosstalk coupling strength is $g_{ab}\approx g(C_c/C_\Sigma ),$
where\textrm{\ }$C_\Sigma =2C_c+C_q$ is the sum of the two coupling
capacitances and qutrit self capacitance. For $C_c\sim $ $1$ fF and $%
C_\Sigma \sim $ $10^2$ fF (the typical values in experiments [54]), we have $%
g_{ab}\leq 0.1g.$ Thus, the condition\textbf{\ }$g_{ab}\leq 2g$\textbf{\ }%
can be easily satisfied\textbf{.}

\section*{Discussion}

We have shown a way to generate the NOON state of two resonators by using a
superconducting coupler transmon qutrit. Unlike the previous schemes, it
requires neither two initially entangled qutrits nor the
photon-number-dependent rotations on the qutrit, and hence is simple, fast,
and robust. Our further numerical simulation shows that a high-fidelity
generation of the NOON state with $N\leq 3$ is feasible within the present
circuit QED techniques. Hence, the present scheme is a significant
development\textbf{\ }for the generation of the NOON state with
superconducting circuit QED, and we hope that the proposed scheme will
stimulate further experimental activities. Finally, it is noted that this
proposal is quite general and can be applied when the coupler qutrit is a
different physical system such as a quantum dot, an NV center, and a
superconducting flux, charge, or phase qutrit.

\section*{Methods}

\noindent \textbf{Hamiltonian and Jaynes-Cumming evolution.} Consider that
resonator $b$ is decoupled from the qutrit, while resonator $a$ is coupled
to the $\left\vert g\right\rangle $ $\leftrightarrow $ $\left\vert
e\right\rangle $ transition of the qutrit but is decoupled from (far
off-resonant with) the $\left\vert e\right\rangle $ $\leftrightarrow $ $%
\left\vert f\right\rangle $ transition [Fig.~2(a)]. In this case, the
Hamiltonian of the whole system in the Schr\"{o}dinger picture is given by $%
H=H_{0}+H_{int},$ with
\begin{eqnarray}
H_{0} &=&E_{g}\left\vert g\right\rangle \left\langle g\right\vert
+E_{e}\left\vert e\right\rangle \left\langle e\right\vert +E_{f}\left\vert
f\right\rangle \left\langle f\right\vert +\hbar \omega _{a}a^{+}a+\hbar
\omega _{b}b^{+}b,  \nonumber \\
H_{int} &=&\hbar \left( g_{eg}a^{+}\left\vert g\right\rangle \left\langle
e\right\vert \right) +h.c.,
\end{eqnarray}
where $\omega _{a}$ ($\omega _{b}$) is the frequency of resonator $a$ ($b$)$%
, $ $H_{0}$ is the free Hamiltonian of the whole system$,$ and $H_{int}$ is
the interaction Hamiltonian between the qutrit and resonator $a.$ In the
interaction picture with respect to the free Hamiltonian $H_{0}$, one can
easily get
\begin{eqnarray}
H_{I} &=&e^{iH_{0}t/\hbar }H_{int}e^{-iH_{0}t/\hbar }  \nonumber \\
&=&\hbar \left[ g_{eg}e^{-i\left( \omega _{eg}-\omega _{a}\right)
t}a^{+}\left\vert g\right\rangle \left\langle e\right\vert \right] +h.c.,
\end{eqnarray}
where $\omega _{eg}=\left( E_{e}-E_{g}\right) /\hbar $ is the transition
frequency between the two levels $\left\vert g\right\rangle $ and $%
\left\vert e\right\rangle $ of the qutrit. In the case when $\omega
_{eg}=\omega _{a},$ i.e., resonator $a$ is resonant with the $\left\vert
g\right\rangle \leftrightarrow \left\vert e\right\rangle $\ transition of
the qutrit, the Hamiltonian (17) becomes $H_{I}=\hbar \left(
g_{eg}a^{+}\left\vert g\right\rangle \left\langle e\right\vert \right)
+h.c., $ which is the Hamiltonian used for the first $N$ steps of the NOON
state preparation. It is easy to show that under this Hamiltonian, the time
evolution of the state $\left\vert e\right\rangle \left\vert n\right\rangle
_{a}$ of the qutrit and the resonator $a$ is described by
\begin{equation}
\left\vert e\right\rangle \left\vert n\right\rangle _{a}\rightarrow \cos (%
\sqrt{n+1}g_{eg}t)\left\vert e\right\rangle \left\vert n\right\rangle
_{a}-i\sin \left( \sqrt{n+1}g_{eg}t\right) \left\vert g\right\rangle
\left\vert n+1\right\rangle _{a},
\end{equation}
where $\left\vert n\right\rangle _{a}$ and $\left\vert n+1\right\rangle _{a}$
are the photon-number states of resonator $a$. Choosing $t=\pi /\left( 2%
\sqrt{n+1}g_{eg}\right) ,$ we obtain the transformation $\left\vert
e\right\rangle \left\vert n\right\rangle _{a}\rightarrow -i\left\vert
g\right\rangle \left\vert n+1\right\rangle _{a},$ which was used for the
first $N$ steps of the NOON state preparation above.

Next, consider that resonator $a$ is decoupled from the qutrit, while
resonator $b$\ is resonant with the $\left| e\right\rangle \leftrightarrow
\left| f\right\rangle $\ transition of the qutrit but is far off-resonant
with the $\left| g\right\rangle \leftrightarrow \left| e\right\rangle $\
transition [Fig.~2(c)]. In this case, the Hamiltonian in the interaction
picture is\textbf{\ }$H_I=\hbar \left( g_{fe}b^{+}\left| e\right\rangle
\left\langle f\right| \right) +h.c.,$ which is the one used for the second $%
N $ steps of the NOON state preparation. It is straightforward to show that
under this Hamiltonian, the time evolution of the state $\left|
f\right\rangle \left| n\right\rangle _b$ of the qutrit and the resonator $b$
is characterized by
\begin{equation}
\left| f\right\rangle \left| n\right\rangle _b\rightarrow \cos (\sqrt{n+1}%
g_{fe}t)\left| f\right\rangle \left| n\right\rangle _b-i\sin \left( \sqrt{n+1%
}g_{fe}t\right) \left| e\right\rangle \left| n+1\right\rangle _b,
\end{equation}
where $\left| n\right\rangle _b$ and $\left| n+1\right\rangle _b$ are the
photon-number states of resonator $b$. For $t=\pi /\left( 2\sqrt{n+1}%
g_{fe}\right) ,$ we have $\left| f\right\rangle \left| n\right\rangle
_b\rightarrow -i\left| e\right\rangle \left| n+1\right\rangle _b,$ which was
used for the second $N$ steps of the NOON state preparation above.

\noindent \textbf{Qutrit-pulse resonant interaction.} When a classical pulse
is resonant with the transition between the level $\left\vert k\right\rangle
$ and the higher-energy level $\left\vert l\right\rangle $ of the qutrit,
the interaction Hamiltonian in the interaction picture is given by $%
H_{I}=\Omega _{lk}e^{i\varphi }\left\vert k\right\rangle \left\langle
l\right\vert +h.c.$. From this Hamiltonian, it is easy to find that a pulse
of duration $t$ results in the following rotation
\begin{equation}
\left\vert k\right\rangle \rightarrow \cos \Omega _{lk}t\left\vert
k\right\rangle -ie^{-i\varphi }\sin \Omega _{lk}t\left\vert l\right\rangle .
\end{equation}
Based on Eq. (20), one can see that when the two levels $\left\vert
k\right\rangle $ and $\left\vert l\right\rangle $ are $\left\vert
g\right\rangle $ and $\left\vert e\right\rangle $ of the qutrit$,$ we have
the transformation $\left\vert g\right\rangle \rightarrow \left\vert
e\right\rangle $ for $\varphi =-\pi /2$ and $t=\pi /\left( 2\Omega
_{eg}\right) ,$ which was used for the first $N-1$ steps and the last step
of the NOON state preparation. On the other hand, Eq. (20) shows that when
the two levels $\left\vert k\right\rangle $ and $\left\vert l\right\rangle $
are $\left\vert e\right\rangle $ and $\left\vert f\right\rangle $ of the
qutrit$,$ we have the transformation $\left\vert e\right\rangle \rightarrow
\left\vert f\right\rangle $ for $\varphi =-\pi /2$ and $t=\pi /\left(
2\Omega _{fe}\right) ,$ which was used for the first $N-1$ of the second $N$
steps of the NOON state preparation.

\begin{addendum}

\item[Acknowledgments]
S.B. Zheng was supported by the Major State Basic Research Development Program of China under Grant No. 2012CB921601.
C.P.Y. was supported in part by the National Natural Science Foundation of China under
Grant Nos. 11074062 and 11374083, the Zhejiang Natural Science Foundation under Grant No.
LZ13A040002, and the funds from Hangzhou Normal University under Grant Nos.
HSQK0081 and PD13002004. Q.P.S. was supported by the National Natural
Science Foundation of China under Grant No. 11147186. This work was also supported
by the funds from Hangzhou City for the Hangzhou-City Quantum Information and Quantum Optics
Innovation Research Team, and the Open Fund from the SKLPS of ECNU.

\item[Author contributions]
Q.P. carried out all calculations under the guidance of S.B. and C.P..
All authors contributed to the interpretation of the work and the writing of the manuscript.

\item[Additional information]
Competing financial interests: The authors declare no competing financial
interests.
\end{addendum}

\clearpage

\textbf{Figure 1:} Setup for two resonators $a$ and $b$ coupled by a
superconducting transmon qutrit. Each resonator here is a one-dimensional
coplanar waveguide transmission line resonator. The circle $A$ represents a
superconducting transmon qutrit, which is capacitively coupled to each
resonator via a capacitance $C_c$. \bigskip

\textbf{Figure 2:} (a) Resonator $a$ is far-off resonant with the $\left|
e\right\rangle \leftrightarrow \left| f\right\rangle $ transition but
resonant with the $\left|g\right\rangle \leftrightarrow \left|
e\right\rangle $ transition. (b) The pulse is far-off resonant with the $%
\left| e\right\rangle \leftrightarrow \left| f\right\rangle $ transition but
resonant with the $\left| g\right\rangle \leftrightarrow \left|
e\right\rangle $ transition. (c) Resonator $b$ is far-off resonant with the $%
\left| g\right\rangle \leftrightarrow \left| e\right\rangle $ transition but
resonant with the $\left|e\right\rangle \leftrightarrow \left|
f\right\rangle $ transition. (d) The pulse is far-off resonant with the $%
\left| g\right\rangle \leftrightarrow \left| e\right\rangle $ transition but
resonant with the $\left| e\right\rangle \leftrightarrow \left|
f\right\rangle $ transition. \bigskip

\textbf{Figure 3:} (a) and (c) Illustration of qutrit-resonator
interactions. (b) and (d) Illustration of qutrit-pulse interactions. In (a),
resonator $a$ is resonant to the $\left| g\right\rangle \leftrightarrow
\left| e\right\rangle $ transition with coupling constant $g_{eg},$ while
off-resonant to the $\left| e\right\rangle \leftrightarrow \left|
f\right\rangle $ transition with coupling constant $\widetilde{g}_{fe}$ and
detuning\textbf{\ }$\delta _1=\omega _{fe}-\omega _a<0;$ resonator $b$ is
off-resonant to the $\left| g\right\rangle \leftrightarrow \left|
e\right\rangle $ transition with coupling constant $\mu _{eg}$ and detuning%
\textbf{\ }$\delta _{eg}=\omega _{eg}-\omega _b>0$, and off-resonant to the $%
\left| e\right\rangle \leftrightarrow \left| f\right\rangle $ transition
with coupling constant $\mu _{fe}$ and detuning\textbf{\ }$\delta
_{fe}=\omega _{fe}-\omega _b>0.$ In (c), resonator $b$ is resonant to the $%
\left| e\right\rangle \leftrightarrow \left| f\right\rangle $ transition
with coupling constant $g_{fe},$ while off-resonant to the $\left|
g\right\rangle \leftrightarrow \left| e\right\rangle $ transition with
coupling constant $\widetilde{g}_{eg} $ and detuning\textbf{\ }$\delta
_3=\omega _{eg}^{\prime }-\omega _b>0;$ resonator $a$ is off-resonant to the
$\left| g\right\rangle \leftrightarrow \left| e\right\rangle $ transition
with coupling constant $\widetilde{\mu }_{eg}$ and detuning\textbf{\ }$%
\widetilde{\delta }_{eg}=\omega _{eg}^{\prime }-\omega _a<0,$ and
off-resonant to the $\left| e\right\rangle \leftrightarrow \left|
f\right\rangle $ transition with coupling constant\textbf{\ }$\widetilde{\mu
}_{fe}$\textbf{\ }and detuning\textbf{\ }$\widetilde{\delta }_{fe}=\omega
_{fe}^{\prime }-\omega _a<0.$ In (b), a pulse (with frequency $\omega
=\omega _{eg}$) is resonant to the $\left| g\right\rangle \leftrightarrow
\left| e\right\rangle $ transition with Rabi frequency $\Omega _{eg},$ but
off-resonant to the $\left| e\right\rangle \leftrightarrow \left|
f\right\rangle $ transition with Rabi frequency $\widetilde{\Omega }_{fe}$
and detuning\textbf{\ }$\delta _2=\omega _{fe}-\omega =\omega _{fe}-\omega
_{eg}<0$. In (d), a pulse (with frequency $\omega =\omega _{fe}^{\prime }$)
is resonant to the $\left| e\right\rangle \leftrightarrow \left|
f\right\rangle $ transition with Rabi frequency $\Omega _{fe},$ but
off-resonant to the $\left| g\right\rangle \leftrightarrow \left|
e\right\rangle $ transition with Rabi frequency $\widetilde{\Omega }_{eg}$
and detuning $\delta _4=\omega _{eg}^{\prime }-\omega =\omega _{eg}^{\prime
}-\omega _{fe}^{\prime }>0.$ The qutrit-resonator interactions during the
pulses of (b) and (d) are the same as those shown in (a) and (c),
respectively, and have been taken into account in the numerical simulation.
Here, $\delta _1=\delta _2$ because of $\omega =\omega _a,$ and $\delta
_3=\delta _4$ due to $\omega =\omega _b$. \bigskip

\textbf{Figure 4:} Fidelity versus $N$. Refer to the text for the parameters
used in the numerical calculation. For $N=1,2,3,4,5$ and $g_{ab}=2g,$ the
fidelities are $\sim $0.947, 0.861, 0.762, 0.669, 0.577, respectively.
\bigskip

\textbf{Figure 5:} Fidelity versus $\{T,g\}$ for $N=3$. The plot was drawn
by setting $\gamma _{\varphi ,f}^{-1}=\gamma _{\varphi ,e}^{-1}=T$, $\gamma
_{fe}^{-1}=T/2$, $\gamma _{eg}^{-1}=T.$ Other parameters used in the
numerical simulation are the same as those used in Fig. 4. \bigskip

\clearpage
\begin{figure}[tbp]
\begin{center}
\epsfig{file=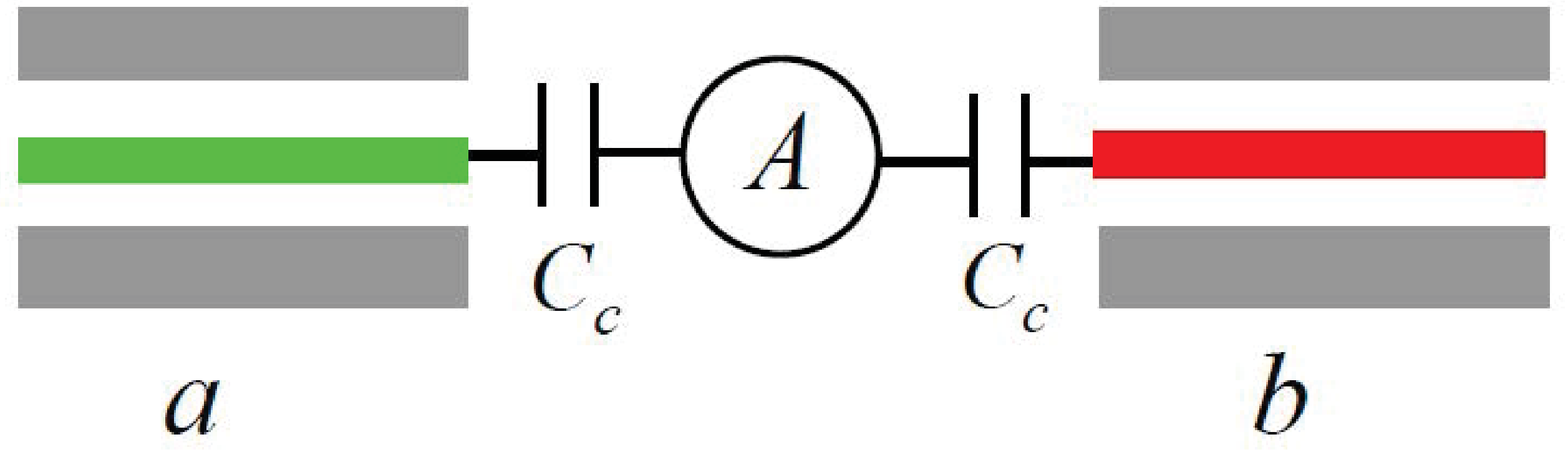,width=15cm}
\end{center}
\caption{}
\label{fig:1}
\end{figure}

\begin{figure}[tbp]
\begin{center}
\epsfig{file=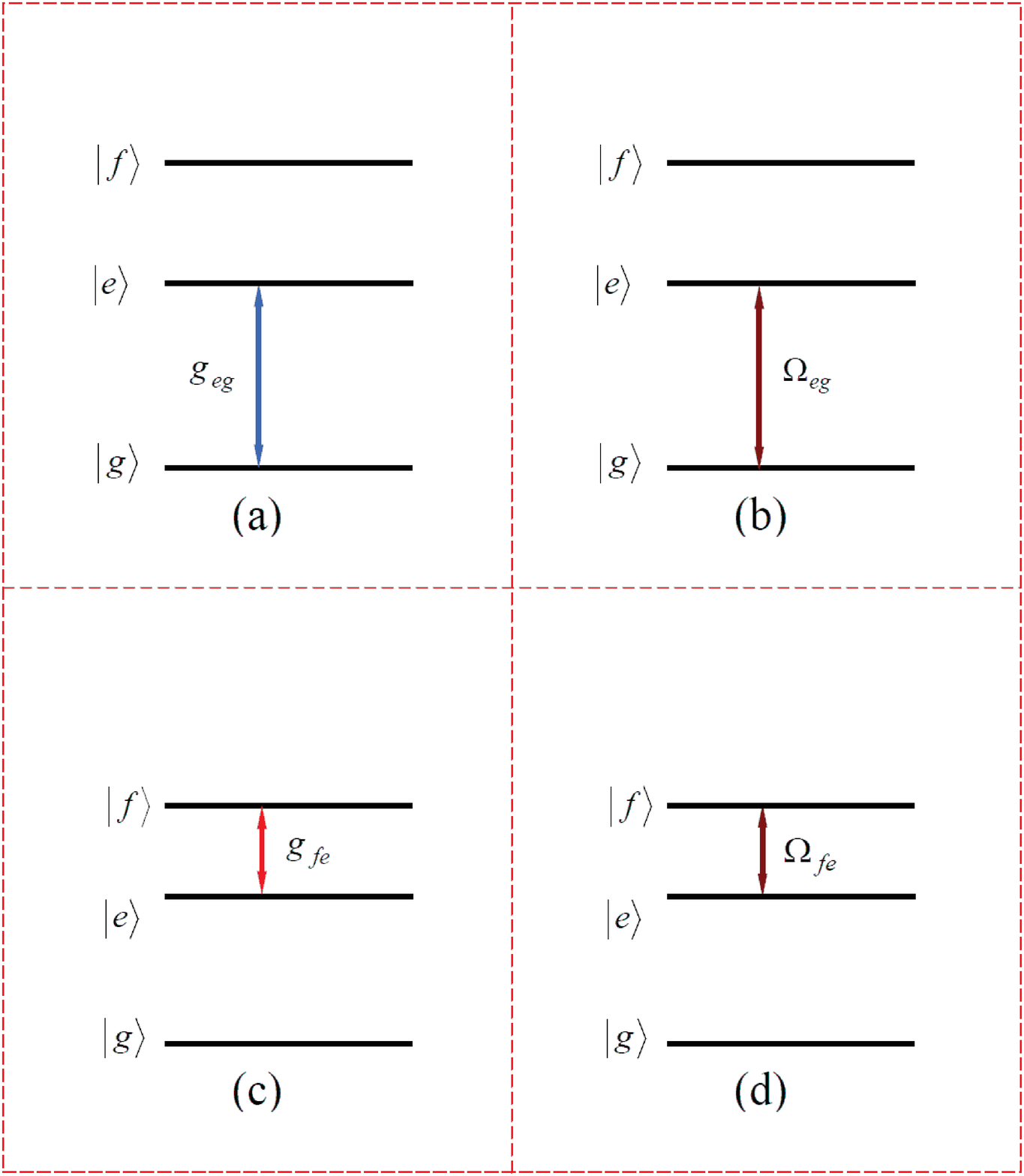,width=13cm}
\end{center}
\caption{}
\label{fig:2}
\end{figure}

\begin{figure}[tbp]
\begin{center}
\epsfig{file=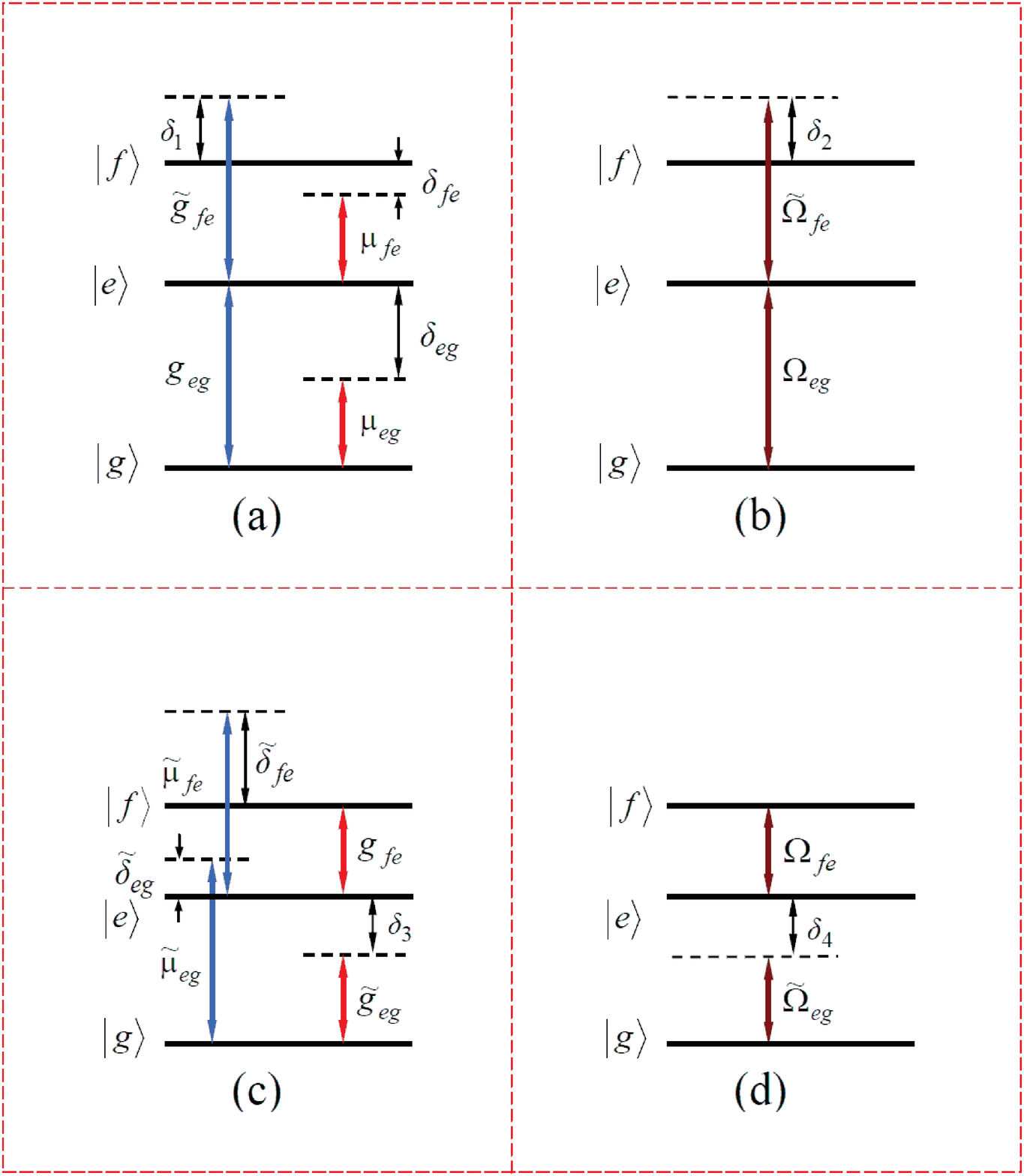,width=13cm}
\end{center}
\caption{}
\label{fig:3}
\end{figure}

\begin{figure}[tbp]
\begin{center}
\epsfig{file=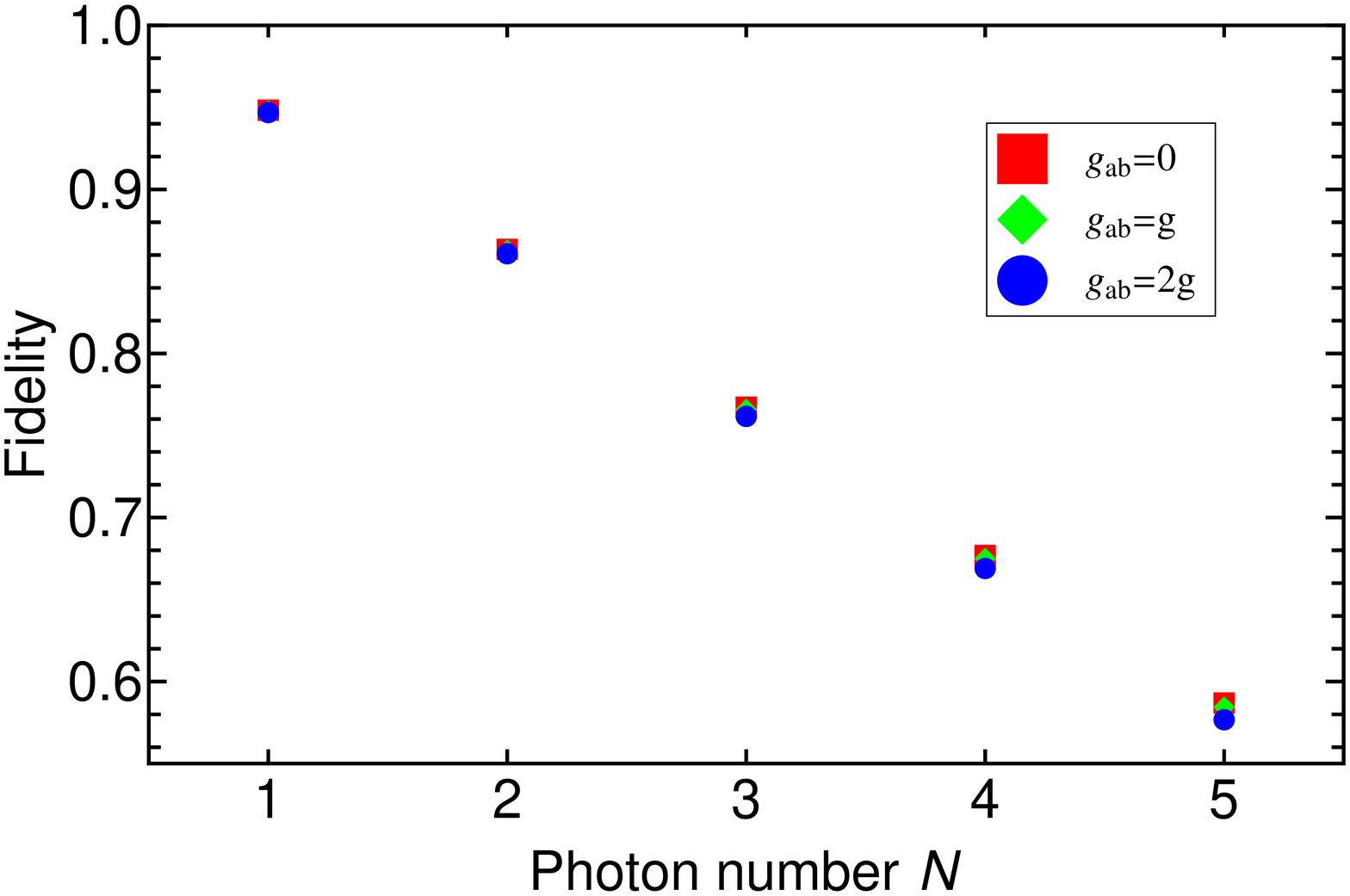,width=15cm}
\end{center}
\caption{}
\label{fig:4}
\end{figure}

\begin{figure}[tbp]
\begin{center}
\epsfig{file=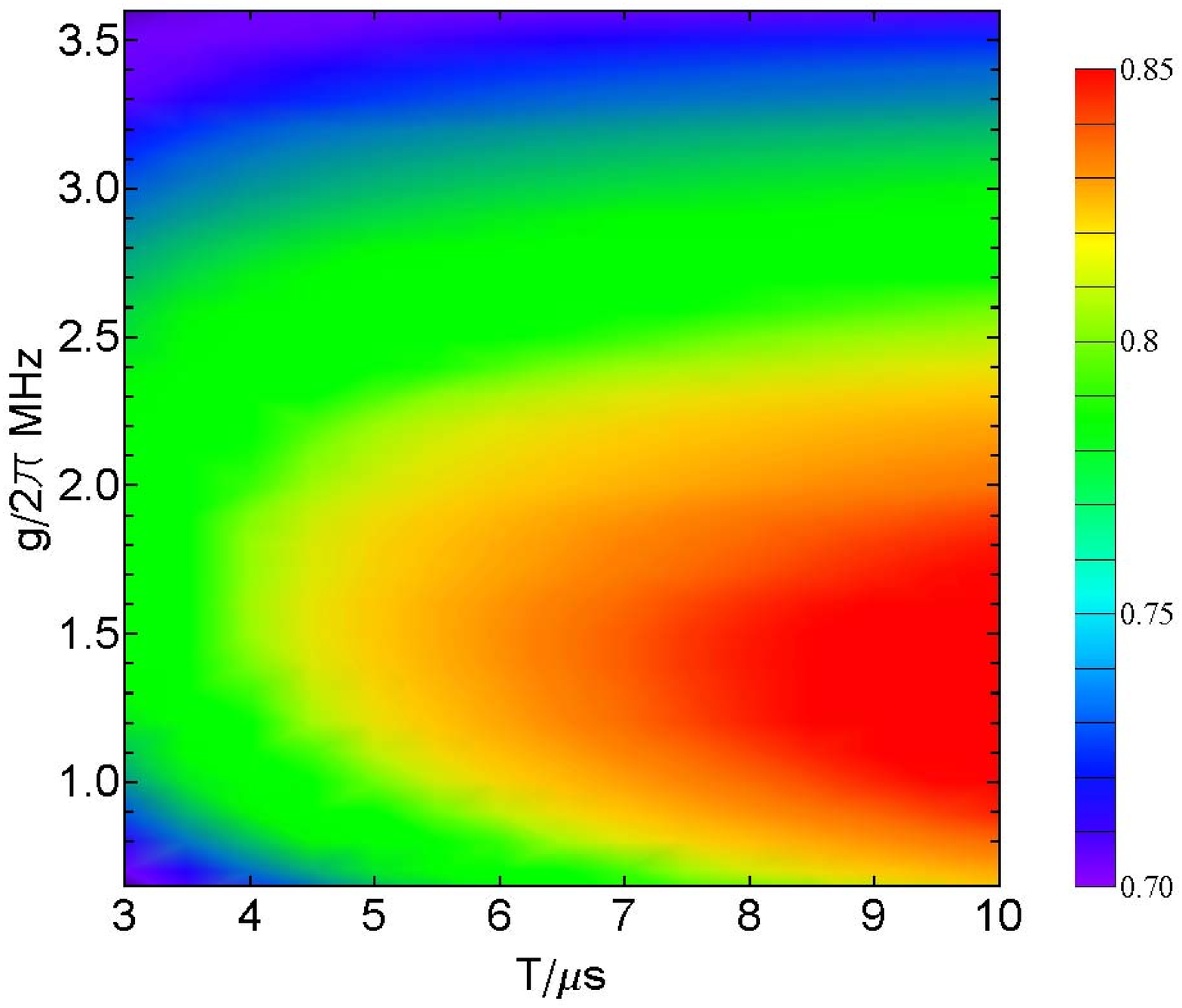,width=13cm}
\end{center}
\caption{}
\label{fig:5}
\end{figure}


\end{document}